\documentclass[
  twocolumn,
  prl,
  amsmath,
  amssymb,
  superscriptaddress,
  floatfix
]{revtex4}

\usepackage{bm}
\usepackage{dsfont}
\usepackage{graphicx}
\usepackage{pifont}
\usepackage{textcomp}
\usepackage{gensymb}
\usepackage{accents}
\usepackage{upgreek}
\usepackage[cmyk,dvipsnames]{xcolor}
\usepackage{color}
\usepackage{ulem}

\usepackage{tabularx}
\newcolumntype{Y}{>{\centering\arraybackslash}X}

\newcommand{\s}{\sum\limits}
\newcommand{\be}{\begin{equation}}
\newcommand{\e}{\end{equation}}
\newcommand{\n}{\nonumber}
\newcommand{\la}{\langle}
\newcommand{\ra}{\rangle}
\newcommand{\bb}{\boldsymbol}
\definecolor{pacificb}{HTML}{1CA9C9}

\makeatletter
\def\NAT@bibsetnum#1{%
 \setlength{\topsep}{\z@}%
 \NATx@bibsetnum{#1}%
}%

\newcommand*{\supplementarystart}{%
  \close@column@grid%
  \clearpage%
  \onecolumngrid%
  \setcounter{enumiv}{0} 
  \setcounter{equation}{0} 
  \setcounter{figure}{0} 
  \setcounter{table}{0} 
  \setcounter{page}{1}
  \c@secnumdepth=4
  \renewcommand{\theequation}{s\arabic{equation}} 
  \renewcommand{\bibnumfmt}[1]{[s##1]} 
  \renewcommand{\@onlinecite}{s\citealp} 
  \renewcommand{\cite}[1]{{[}\onlinecite{##1}{]}}
  \renewcommand{\thefigure}{s\arabic{figure}}
  \renewcommand{\thetable}{s\Roman{table}}
  \renewcommand{\thepage}{s\arabic{page}}
}
\makeatother

\begin{document}

\title{Stability and lifetime of antiferromagnetic skyrmions}

\author{P.\,F.\,Bessarab}
\affiliation{Science Institute of the University of Iceland, IS-107 Reykjav\'ik, Iceland}
\affiliation{Institute of Theoretical Physics and Astrophysics, University of Kiel, Leibnizstrasse 15, 24098 Kiel, Germany}
\affiliation{ITMO University, Saint Petersburg 197101, Russia}
\author{D.\,Yudin}
\affiliation{Deep Quantum Labs, Skolkovo Institute of Science and Technology, Moscow 121205, Russia}
\affiliation{ITMO University, Saint Petersburg 197101, Russia}
\author{D.\,R.\,Gulevich}
\affiliation{ITMO University, Saint Petersburg 197101, Russia}
\author{P.\,Wadley}
\affiliation{School of Physics and Astronomy, University of Nottingham, Nottingham NG7 2RD, United Kingdom}
\author{M.\,Titov}
\affiliation{Radboud University Nijmegen, Institute for Molecules and Materials, NL-6525 AJ Nijmegen, The Netherlands}
\affiliation{ITMO University, Saint Petersburg 197101, Russia}
\author{Oleg A.\,Tretiakov}
\email{o.tretiakov@unsw.edu.au}
\affiliation{School of Physics, The University of New South Wales, Sydney 2052, Australia}
\affiliation{National University of Science and Technology MISiS, Moscow 119049, Russia}

\begin{abstract}
The two-dimensional Heisenberg exchange model with out-of-plane anisotropy and Dzyaloshinskii-Moriya interaction is employed to investigate the lifetime and stability of antiferromagnetic (AFM) skyrmion as a function of temperature and external magnetic field. An isolated AFM skyrmion is metastable at zero temperature in a certain parameter range set by two boundaries separating the skyrmion state from the uniform AFM phase and a stripe domain phase. The distribution of the energy barriers for the AFM skyrmion decay into the uniform AFM state complements the zero-temperature stability diagram and demonstrates that the skyrmion stability region is significantly narrowed at finite temperatures. We show that the AFM skyrmion stability can be enhanced by an application of magnetic field, whose strength is comparable with the spin-flop field. This stabilization of AFM skyrmions in external magnetic fields is in sharp contrast to the behavior of their ferromagnetic counterparts. Furthermore, we demonstrate that the AFM skyrmions are stable on the timescales of milliseconds below 50 K for realistic material parameters, making it feasible to observe them in modern experiments.
\end{abstract}

\maketitle

{\it Introduction.} Localized topological spin-textures, such as magnetic skyrmions~\cite{Bogdanov1989}, hold great promise as a basis for future digital technologies~\cite{Yu2010, Yu2011,Heinze2011,Romming2013, Rosch2013, Leonov2016, Moreau2016,Soumy2017,Dupe2016,Garst2016,Rosch2017,Shen2019}. Information flow in next-generation spintronic devices could be associated with metastable isolated skyrmions guided along magnetic strips~\cite{Fert2013,Tomasello2014,Koshibae2015,Kang2016}. Such skyrmion racetrack schemes are expected to considerably reduce the power consumption of data processing due to the sensitivity of skyrmions to external stimuli, particularly electric current~\cite{Nagaosa2013, Hoffmann2015, Mochizuki2015, Woo2016, Lucia2017,Abanov2017,Woo2017}. However, isolated skyrmions in chiral ferromagnets suffer from the skyrmion Hall effect~\cite{Litzius2017,Hoffmann2016_Sk}, which potentially limits the use of skyrmions for racetrack nanodevices. The Skyrmion Hall effect may be understood using collective coordinate approach to topological spin textures~\cite{Thiele73,Tretiakov08,Clarke08}, where it translates into a generalized gyrotropic (Magnus) force~\cite{Everschor2011,Iwasaki2013,Tomasello2014, Ado2017} acting on a skyrmion in a direction that is transverse to the applied electric current direction,  and thus eventually pushing it over the edge of the nanotrack.

Recently, it has been suggested based on both analytical arguments and micromagnetic simulations that unfavorable effect of the topological Magnus force on skyrmions can completely cancel out in chiral antiferromagnetic (AFM) materials~\cite{Barker2016,Zhang2016a,Zhang2016b}. In such AFM skyrmions, the Magnus force on one magnetic sublattice is equal in magnitude but has an opposite sign to the one on the other sublattice, thus leading to straight skyrmion trajectories along the applied current and furthermore greatly enhanced velocities compared to its FM counterpart~\cite{Barker2016,Zhang2016a,Zhang2016b,Velkov2016}. Additionally, using micromagnetic simulations it has been proposed  how to create the AFM skyrmions by injecting vertically spin-polarized current into a nanodisk with a uniform AFM state \cite{Zhang2016a}. A possible experimental realization of an isolated skyrmion as well as a skyrmion lattice has been suggested by using a standard bipartite lattice in which each sublattice supports a skyrmion crystal (e.g, honeycomb lattice) coupled to an AFM \cite{Gobel2017}. Moreover, the topological spin Hall effect has been studied in AFM skyrmions and its impact on the current-induced motion has been demonstrated \cite{Mokrousov2017, Akosa2017}.

Although there has been an enormous progress in studying the dynamics of AFM spin textures~\cite{Brataas2011,Tveten2013,Gomonay2016, Rodrigues2017} and AFM materials in general \cite{Jungwirth2016,Wadley2016}, the AFM skyrmions have not been experimentally discovered yet. This may have to do with the overall challenge in detection of N\'eel-order spin textures~\cite{MacDonald2011},  as well as finding chiral AFM material with the appropriate parameters~\cite{Barker2016}. The stability of AFM skyrmions could also be an issue. In continuum models skyrmion annihilation into a uniform AFM state is strictly prohibited due to different topological charges for N\'eel order parameter of the target states. For physical systems with magnetic moments localized on atoms, topological protection is not strict, which translates into finite energy barriers separating skyrmions from topologically distinct states. Thermal fluctuations can bring the system over the barrier and spontaneously destroy the skyrmoin state, resulting in a finite skyrmion lifetime at nonzero temperature. If the lifetime is too short on a scale of available experimental techniques, such as spin-polarized scanning-tunneling microscopy (SP-STM) or magnetic exchange force microscopy~\cite{Wiesendanger2009}, the AFM skyrmion would decay before being detected. A long enough lifetime is an essential prerequisite for the use of skyrmions in applications.

In this Rapid Communication, we analyze stability of AFM skyrmions. Both the activation energy for the skyrmion decay and the skyrmion lifetime are evaluated as functions of material parameters, temperature, and magnetic field using harmonic transition state theory for spins~\cite{Bessarab2012}. This analysis makes it possible to quantify the skyrmion stability at macroscopic time scales. We complement the zero-temperature phase diagram for an isolated AFM skyrmion with the distribution of energy barriers for the skyrmion collapse into the uniform AFM phase. Our analysis demonstrates that the stability region may be significantly narrowed even at small temperatures. However, the AFM skyrmions can be further stabilized by a magnetic field, which is in sharp contrast to their FM counterparts. The AFM skyrmions are shown to be rather stable at  50 K and below for typical AFMs, where they may be observed using modern techniques for the detection of N\'eel order parameter~\cite{MacDonald2011}. 

{\it Methods.} We study a monolayer AFM spin system on a square lattice using localized-moment Hamiltonian equipped with Heisenberg exchange coupling, antisymmetric Dzyaloshinskii-Moriya interaction (DMI), out-of-plane anisotropy, and Zeeman term. The energy functional reads
\begin{align}\n
E=&\frac{J}{2}\s_{\la i,j\ra} \bb{m}_i\cdot\bb{m}_j -\frac{D}{2}\s_{\la i,j\ra} \bb{d}_{ij} \cdot (\bb{m}_i\times\bb{m}_j)\\
& - K\s_{i} (m_i^z)^2- MB \s_i m^z_i,
\label{E}
\end{align}
where $\la i, j\ra$ denotes the summation over the nearest neighbors, $\bb{m}_i$ is the unit vector in the direction of the magnetic moment on site $i$,  $J$ and $D$ are the exchange and DMI constants, respectively, $K$ is the anisotropy constant, $B$ is the magnetic field, and $M$ is the magnitude of the on-site magnetic moment. Both the anisotropy and external field are perpendicular to the AFM film. The DMI unit vectors $\bb{d}_{ij}$ lie in the film plane and point perpendicular to the bond connecting sites $i$ and $j$. Dipolar interactions are not included in the Hamiltonian because their effect is suppressed by AFM texture. Equation~(\ref{E}) defines a multidimensional energy surface as a function of the orientation of magnetic moments, where in a certain parameter range the local minima corresponding to N\'eel-type skyrmions emerge~\cite{Barker2016}. We obtain a skyrmion solution by taking a rough initial guess for the skyrmion profile and relaxing it to a local energy minimum. We place only one single skyrmion in the simulated system and apply periodic boundary conditions to model an extended two-dimensional system. We define the skyrmion radius as a square root of the area enclosed within the $m_z =0$ contour divided by $\pi$. The computational domain is chosen to be large enough for an isolated equilibrium skyrmion not to be affected by the boundaries.

The lifetime of AFM skyrmions, $\tau$, is calculated using the harmonic transition state theory for magnetic systems~\cite{Bessarab2012}. Similar approaches are employed in various branches of condensed matter physics for the evaluation of the decay rate of a metastable state~\cite{Tretiakov2003,Tretiakov2005}. The theory predicts an Arrhenius expression for the lifetime as a function of temperature $T$,
\be
\label{tau}
\tau(T)=\tau^\prime\,e^{\Delta/k_{\text{B}}T}.
\e
Here the activation energy $\Delta$ is given by the energy difference between the skyrmion-state local minimum and relevant saddle point located on the minimum energy path connecting the skyrmion configuration with the uniform AFM phase. The preexponential factor $\tau^\prime$ is defined by the curvature of the energy surface at the saddle point and at the skyrmion-state minimum. It could acquire a power-law temperature dependence due to soft modes corresponding to the translational motion of the skyrmion structure~\cite{Braun1994,Bessarab2018,suppl}. The identification of minimum energy paths and the corresponding saddle points on the energy surface is carried out using the geodesic nudged elastic band (GNEB) method~\cite{Bessarab2015}. GNEB calculations have previously been used to identify mechanisms and energy barriers for the skyrmion annihilation in FM materials~\cite{rohart2016,bessarab2017,Lobanov2016,Stosic2017,cortes2017,Uzdin2018}. Here, we only consider the minimum energy paths that correspond to the radial collapse of the AFM skyrmion.

\begin{figure}[tbh!]
\begin{center}
\includegraphics[width=3.5in]{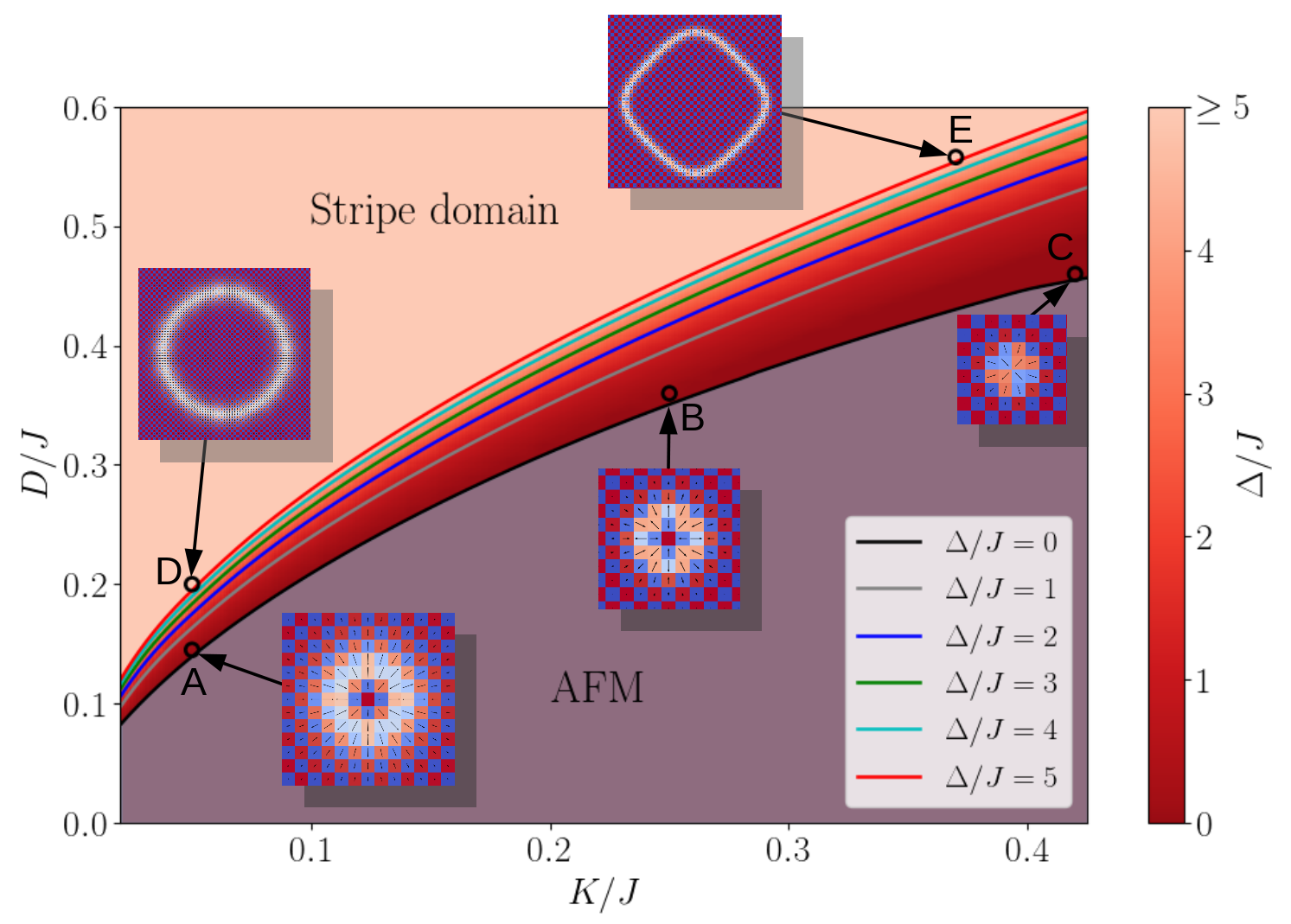}
\caption{\label{fig:stability} 
 AFM skyrmion stability diagram at zero temperature and magnetic field. Isolated skyrmions exist in the sector between the dashed and solid black lines. At the lower boundary (black solid line) skyrmions spontaneously collapse into the uniform AFM phase, while at the upper boundary (dashed black line) they strip out into stripe domains. The strip-out boundary obtained in the continuous N\'eel vector model is shown by the dotted line. The height of energy barriers $\Delta$ for the skyrmion decay into the uniform AFM state is represented by the red color intensity. Several contours of constant energy barriers are depicted by colored solid lines. The insets show spin configurations for several points on the stability diagram.
}
\end{center}
\end{figure}

{\it Stability diagram.} The zero-temperature AFM skyrmion stability diagram for a monolayer AFM  in the absence of magnetic field is presented in Fig.~\ref{fig:stability}. The sector, where isolated skyrmions exist as metastable states in the uniform AFM background (it was demonstrated for ferromagnetic skyrmions in Refs.~\cite{wilson2014,leonov2016a}), is situated between the uniform AFM state from below and stripe AFM domain from above.  At the lower boundary of this sector (black solid line in Fig.~\ref{fig:stability}), the energy barrier $\Delta$ vanishes and skyrmions decay into the uniform AFM state. Note that the skyrmions collapse with finite radii, which is a consequence of the discreteness of the model used here (see insets A, B, and C in Fig.~\ref{fig:stability}). At the upper boundary (dashed line in Fig.~\ref{fig:stability}), isolated skyrmions expand to become stripe domains~\cite{wilson2014,leonov2016a}, since it is energetically favorable to increase the length of the domain wall separating up and down domains indefinitely. The upper boundary can be obtained analytically within the micromagnetic continuous model for the N\'eel vector.  We can separate two contributions to the energy of a skyrmion of radius $R$: the domain wall energy $E_{dw}(R)$ defined as the energy of the wall of length $2\pi R$ (relative to the AFM phase) and the energy of skyrmion's domain wall curvature defined as the difference $E_c(R) = E_{sk}(R)-E_{dw}(R)>0$, where $E_{sk} (R)$ is the exact energy of a skyrmion. While it is evident that $E_{dw}(R)$ grows linearly with $R$, the curvature energy $E_c(R)$ decreases with $R$ which guarantees the existence of a local energy minimum. As $E_{dw}(R)$ decreases to zero, so does $E_c (R)$. In an infinite system, the skyrmion expands indefinitely at $E_{dw}(R)=0$. According to Ref.~\onlinecite{Bogdanov2002}, this equation is satisfied at
\begin{equation}
\label{criticalD}
D_c(K) = \frac{4}{\pi}\sqrt{\frac{K J}{2}},
\end{equation}
giving the upper critical bound on $D$ for the skyrmion existence, i.e. the strip-out boundary for the continuous N\'eel vector model. It is shown in Fig.~\ref{fig:stability} by the dotted line. The insets in Fig.~\ref{fig:stability} show spin configurations for several points on the diagram, demonstrating that skyrmions tend to become larger in the vicinity of the upper boundary.

Within the AFM skyrmion stability region, as the anisotropy increases, the isolated skyrmions assume the structure of magnetic bubbles where the core with almost uniformly antiparallel N\'eel vector is separated from the AFM background by a domain wall~\cite{Bogdanov1994}. The width of the wall decreases with the anisotropy parameter and at a certain point becomes comparable to the lattice constant of the system. At this point, the orientation of the domain wall becomes affected by the lattice symmetry. In particular, the domain wall tends to propagate along the diagonal directions of the square lattice, which results in equilibrium skyrmionic structures with broken axial symmetry (see inset E of Fig.~\ref{fig:stability}). As expected, this anisotropic effect is less pronounced for the systems, which are well described by the continuum models, i.e., those characterized by small angles between N\'eel vectors at neighboring lattice sites (see inset D in Fig.~\ref{fig:stability}).

We further analyze the distribution of energy barriers $\Delta$ that have to be overcome by an isolated AFM skyrmion to decay into the uniform AFM state (see Fig.~\ref{fig:stability}). As expected, the barrier height increases monotonically as one moves from the lower stability boundary to the upper one, however, the rate of this increase is not constant. In particular, the barrier demonstrates weak dependence on the material parameters in the wide region close to the lower boundary, where the barrier is rather small. The dependences on $K$ and $D$ become more pronounced as one approaches the upper stability line: the barrier increases rapidly, enhancing the stability of large skyrmions. These results suggest that even at low temperatures sufficiently small skyrmions may be easily destroyed by thermal fluctuations in a large lower portion of the stability diagram, thus significantly reducing the AFM skyrmion stability region at finite temperatures. Given the exponential dependence of the lifetime on the energy barrier, it is expected that AFM skyrmions are stable at long time scales in the region close to the upper stability boundary. Such AFM skyrmions may indeed be detected on the experimentally relevant time scales.

\begin{figure}[tbh!]
\centerline{
\includegraphics[width=1.0\columnwidth]{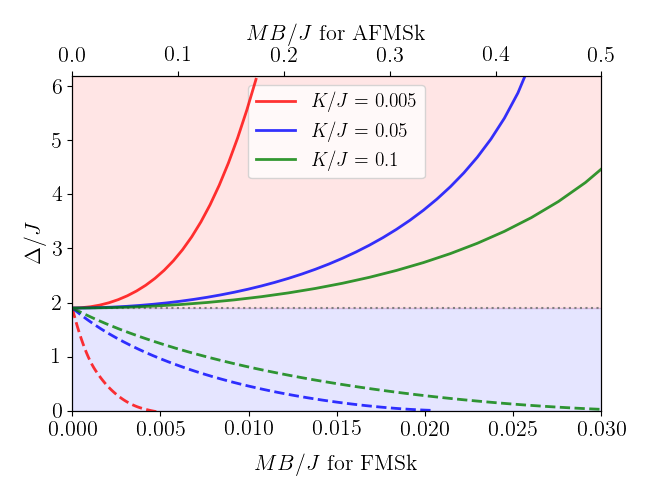}}
\caption{Activation energy $\Delta$ as a function of magnetic field $B$ for AFM skyrmions (AFMSk) shown with solid lines on the light pink background and FM skyrmions (FMSk) represented by dashed lines on the light blue background for several values of anisotropy constant $K$. Note the drastic difference in the magnetic field ranges for the AFMSk and FMSk, respectively. For each value of $K$, the DMI constant was chosen so that the activation energies for the skyrmion decay coincide at zero field.}
\label{fig:Hfield}
\end{figure}

A magnetic field has a nontrivial effect on the AFM skyrmion stability. Figure~\ref{fig:Hfield} shows the barrier for the skyrmion decay into the uniform AFM state as a function of applied field strength. These results are in sharp contrast with the field dependence of the FM skyrmion ($J<0$), where the barrier quickly decreases with the field~\cite{vonMalottki2017, Bessarab2018}. On the contrary, for the AFM skyrmion ($J>0$) the energy barrier is significantly enhanced, but the effect manifests itself at much larger fields.
 
To gain a further insight into this unusual behavior of AFM skyrmions, it is instructive to estimate the skyrmion radius by minimizing the energy functional in the presence of magnetic field. Such an estimate can be obtained analytically within the micromagnetic continuum model for the N\'eel vector~\cite{Keesman2016}, using a trial skyrmion solution for the polar angle $\theta(r)=\pi(1-r/R)$ with $0<r<R$, where the parameter $R$ is associated with the skyrmion radius~\cite{Bogdanov1994}. This minimization analysis gives the radius $R_0=\pi JD/(JK-M^2 B^2/16)$, thus showing that the AFM skyrmion size increases with the field. It is consistent with the field dependence of the energy barrier observed in the numerical simulations, since larger skyrmions correspond to larger energy barriers, as discussed above. One can arrive at the same conclusion from another perspective. It has been shown in Ref.~\cite{Keesman2016} that in the continuous model the energy functional of the AFM skyrmion system at an arbitrary applied field is equivalent to that of the FM counterpart at zero field. Then one can show that the anisotropy constant is  renormalized as follows:
\begin{equation}
\label{renormK}
K^\prime=K-M^2 B^2/(16J).
\end{equation}
Therefore, the enhancement of the energy barrier for the AFM skyrmions in magnetic field can be understood by the effective decrease of the anisotropy in the FM-skyrmion decay problem. Indeed, the decrease in the energy barrier for the FM skyrmion decay with the anisotropy strength has recently been confirmed in Ref.~\cite{Stosic2017} and is evident from our diagram in Fig.~\ref{fig:Hfield}.

{\it AFM skyrmion lifetime.} According to Arrhenius law [see Eq.~\eqref{tau}], the skyrmion lifetime depends exponentially on the energy barrier $\Delta$. However, it is the prefactor $\tau^\prime$ that establishes the time scale. It has to be evaluated for each set of the material parameters, magnetic field, and temperature for a definite identification of the lifetime. The Arrhenius prefactor incorporates both the entropic and dynamical contributions to the skyrmion's stability.

Based on the harmonic transition state theory, we evaluate $\tau^\prime$ as a function of temperature and DMI parameter for a fixed value of the anisotropy constant $K=0.1 J$ and zero magnetic field~\cite{suppl}. We find that the prefactor is temperature independent in the range from $k_B T=0.1 J$ to $k_B T=1 J$, but increases dramatically from 0.4$\times 10^2 \tau_{0}$ to 0.7$\times 10^{14} \tau_{0}$, as $D$ changes from 0.21 to 0.28~\cite{suppl}, which roughly corresponds to the lower and upper boundaries of skyrmion stability for $K=0.1 J$ (see Fig.~\ref{fig:stability}). Here the prefactor is given in units of an intrinsic precession time $\tau_0=M/J\gamma$ with $\gamma$ being the gyromagnetic ratio. Our results demonstrate the importance of definite evaluation of the Arrhenius prefactor for the AFM skyrmionic systems. The assumption that the prefactor does not change under the variation of material parameters would produce inaccurate results concerning skyrmion stability at finite temperatures. A dramatic change in the Arrhenius pre-exponential factor with applied field has recently been observed experimentally for skyrmions in Fe$_{1-x}$Co$_x$Si systems~\cite{Wild2017}.

\begin{figure}[tbh!]
	\centerline{
		\includegraphics[width=1.0\columnwidth]{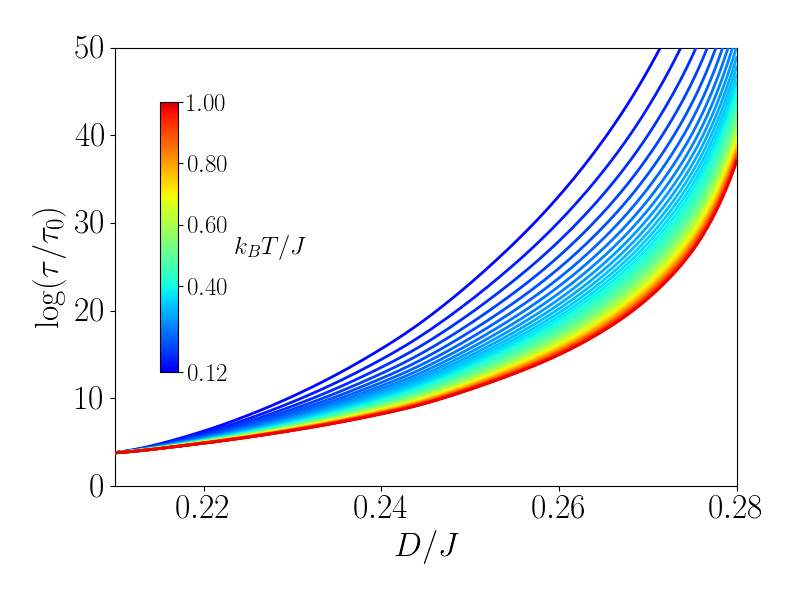}}
	\caption{Lifetime of AFM skyrmion as a function of DMI strength $D$ for several values of temperature $T$ and fixed anisotropy strength $K=0.1J$.}
	\label{fig:lifetime}
\end{figure}

Our results for the skyrmion lifetime are presented in Fig.~\ref{fig:lifetime}  as a function of the DMI parameter for several values of temperature and fixed anisotropy constant $K=0.1 J$. Apart from the exponential decrease of the lifetime with temperature, the plot demonstrates a sharp dependence of the skyrmion lifetime on the DMI parameter. Overall, the AFM skyrmions become more stable as $D$ increases. We point out that this stabilization of skyrmions occurs due to increase of both the energy barrier and the pre-exponential factor. The lifetime is given in units of intrinsic precession time $\tau_0$ and can be estimated for concrete material parameters. By taking the parameters similar to those used in Ref.~\cite{Barker2016}, $J = -9.2\times 10^{-22}$\,J, $D = 5.5\times 10^{-23}$\,J, and $K = 4.6\times 10^{-24}$\,J, one deduces that AFM skyrmions may be stable on the timescales of seconds at temperatures 25 -- 30 K (or milliseconds for temperatures in the range of 50 -- 65 K), and therefore can in principle be detected with SP-STM technique.

{\it Conclusions.} 
We have explored the stability and lifetimes of AFM skyrmions at finite temperatures within harmonic transition state theory formalism. The nonuniform distribution of energy barriers for the skyrmion decay has been shown to lead to a significant reduction of the AFM skyrmion stability region at finite temperatures. Surprisingly, in sharp contrast to FM skyrmions, which rapidly become unstable with increasing magnetic field, its AFM counterparts demonstrate a higher stability in finite magnetic fields. These fields may be large for usual AFMs (corresponding to typical spin-flop fields of $\sim 10$ Tesla~\cite{Shapira1970, Rezende2017}), however the critical fields above which the AFM skyrmion becomes more stable are rather easily achieved in AFMs with a weak AFM exchange.  Furthermore, we have calculated the AFM skyrmion lifetimes to be in the range of milliseconds for a reasonable temperature range (50 -- 65 K), thus demonstrating that AFM skyrmions can be experimentally observed and employed in spintronic applications. We demonstrate that this temperature range can be further increased by applying a magnetic field, as it renormalizes favorably anisotropy [see Eq.~(\ref{renormK})].

{\it Acknowledgements.}
We are grateful to V.\,M. Uzdin and H. J\'onsson for helpful discussions. We would like to thank Ya.\,V. Zhumagulov for confirming the stability diagram in Fig.~\ref{fig:stability} using an alternative method (Monte Carlo simulations presented in the Supplemental Material). P.\,F.\,B. acknowledges support from the Icelandic Research Fund (Grant No. 163048-052), the University of Iceland Research Fund, and Alexander von Humboldt Foundation. P.\,B., D.\,Y., D.\,R.\,G., and M.\,T. acknowledge support from the Russian Science Foundation under Project No.\,17-12-01359. The work of M.\,T. was partially supported by ICC-IMR, Tohoku University (Japan). O.\,A.\,T. acknowledges support by a grant of the Center for Science and Innovation in Spintronics (Core Research Cluster), Tohoku University, by the Ministry of Science and Higher Education of the Russian Federation in the framework of Increase Competitiveness Program of  NUST ``MISiS'' (No.\,K2-2019-006), implemented by a governmental decree dated 16th of March 2013, N 211, and by JSPS and RFBR under the Japan-Russia Research Cooperative Program. 
\bibliography{bibliography}

\end{document}